\documentclass{elsart}

\usepackage[square]{natbib}
 \usepackage{graphics}
 \usepackage{graphicx}
 \usepackage{epsfig}

\usepackage{amssymb}
\begin{document}
\begin{frontmatter}

 \title{The shell model approach\\  to  the rotating turbulence
}

\author{M.~Reshetnyak$^{a,\,b,\, *}$,\,\,} 
\author{B.~Steffen$^{c}$}

  \address{Institute of the Physics of the Earth, Russian Acad.~Sci,
    123995 Moscow, Russia$^a$}
\address{Research Computing Center of Moscow State University,\\ 119899,
Moscow, Russia$^b$}
 \address{Central Institute for Applied Mathematics (ZAM)\\
of Forshungszentrum J$\rm\ddot{u}$ulich, Germany$^c$}
\corauth[cor1]{Corresponding authour. \\ \,\,\, E-mail address:
maxim@uipe-ras.scgis.ru (M.~Reshetnyak).}

\begin{abstract}
Applications of the shell model of turbulence to the case of
rapidly rotating bodies  are considered. Starting from the
classical GOY model we introduce  the Coriolis force and obtain a
$\sim k^{-2}$ spectrum for 3D hydrodynamical turbulence for the
free decay regime as well for the regime with external forcing.
 Additional
modifications of the GOY model providing a realistic form of the
helicity are proposed.
\end{abstract}
\end{frontmatter}

\section{Introduction}
The influence of rotation on the properties of the hydrodynamical
turbulence is of the great importance. This problem appears in the
various geophysical and astrophysical applications \citep{Tabel}
and it requires special treatment. So far the direct 3D numerical
simulations, where implementation of the Coriolis force is
trivial,   cannot
 provide a long enough inertial range of spectrum to reveal
its scaling laws \cite{Frisch},
thus different approaches are needed.

Though the possibility  of
transition from 3D-turbulence state to the
 two-dimensi\-on\-al turbulence due to rotation was
  already predicted by Batchelor   many
years ago \cite{Batch}, only now is the
qualitative description of this processes developing \cite{Zeman}, \cite{Zhou}.
 Having in mind the similar situation in the MHD turbulence \cite{Kraich},
  where the magnetic field plays
 the same role as the rotation in reducing the dimension of the problem, the authors
proposed that the Coriolis force introduces a new characteristic time. This time
 should be  used
instead of the characteristic turn-over time based on the  Kolmogorov's
estimate. Then, instead of Kolmogorov's $-5/3$ slope the spectrum will have a
$-2$ slope.
It appears that this approach is in  agreement
with the direct numerical calculations \cite{Hossain} and
the experiments \cite{Jac}.
 The formal simplicity of this phenomenological approach
   attracts us to implement it in the more
complicated models of turbulence and to check the results for  self-consistence.
For this aim we use the well-known homogeneous isotropical
  GOY shell model  (see overview in \cite{BJPV98})
 and modify it to the case of rotation. The proposed model is tested for the
  regimes of the
free-decay turbulence and for the regime with external forcing. We also
consider  a situation where a non-zero average helicity is generated.
This regime finds its application in
the mean-field dynamo problems.

\section{Basic equations}\label{sec:Largescale}
The evolution of conductive incompressible liquid ($\nabla \cdot {\bf V} =0$),
rotating with the angular velocity
$\Omega$,
 can be described by
the
 Navier-Stokes equation:
  \begin{equation}R_o\left( {\partial{\bf
V}\over\partial t}+\left({\bf V}\cdot\nabla\right){\bf V}\right) =
-\nabla P+ {\bf F}+ E\nabla^2{\bf V}. \label{Nav}
\end{equation}
Eq.(\ref{Nav}) is scaled with the large scale of the body $L$;
velocity $V$,  pressure $P$ and time $t$ are measured in
$\eta/L$,
$\rho\eta^2/L^2$ and $L^2/\eta$,  where $\eta$ is a
 magnetic diffusion, $\rho$ is density,
 $R_o = \eta/2\Omega L^2$ and
 $E = \nu/2\Omega L^2$ are  the Rossby and  Ekman numbers, respectively  and
$\nu$ is the kinematic viscosity. (This kind of scaling takes  its origin
 in  geodyamo problems \cite{Jones}.)
 The force ${\bf F}={\bf F_c}+{\bf f}$ includes the
Coriolis effect:
\begin{equation}
{\bf F_c}= -{\bf 1}_z\times{\bf V} \label{force}
\end{equation}
and a prescribed part $\bf f$. Here
  $\bf 1_z$ is the unit vector along the axis of rotation.

It is worth to note that, in the limit of the  vanishing
viscosity and absence of the external force (${\bf f}=0$),
 equation (\ref{Nav}) conserves kinetic
energy $E_k={V^2/2}$. For the two-dimensional flow the
other integral of motion is the enstrophy
 $E_\Omega=(\nabla\times {\bf
V})^2$ and for  3D case the hydrodynamical helicity
$H={\bf V}\cdot \left(\nabla\times{\bf V}\right)$ is conserved
(see for details, e.g., \cite{BJPV98, Tabel}).

\section{Shell model equations}
\label{sec:Shell}
The idea of the shell model approach  is to mimic the
original  equation (\ref{Nav})  by a dynamical
system with $n_{max}$ complex variables $u_1, u_2,\dots, u_{n_{max}}$ which
represent the
typical magnitude of the velocity  on a certain
length scale. The Fourier space is divided into $n_{max}$ shells
and each shell $k_n$ consists of a set of wave vectors $\bf k$
such that $k_0 2^n <|{\bf k}|<k_0 2^{n+1}$, and thus each shell
corresponds to an octave of the wave numbers. Variable $u_n$ is
the velocity difference over length $\sim 1/k_n$, so that there is
one degree of freedom per shell. The coupling between shells is
considered to preserve the main symmetries and properties of  equation (\ref{Nav}).
  Of course, the shell model is a simplified form of
the description of the turbulence, nevertheless, it is a
reasonable tool in the studies of turbulence. Here and after we
will refer to the GOY model  \cite{BJPV98},  which has the form:
\begin{equation}\begin{array}{l}
R_o {d u_{\rm n}\over dt} =
 R_o i k_{\rm n} \Bigl(u_{n+1}^*u_{\rm n+2}^* -{\epsilon\over
2}u_{\rm n-1}^*u_{\rm n+1}^*-{1-\epsilon\over 4}u_{\rm n-2}^*u_{\rm n-1}^*\Bigr)-
\\ \qquad\qquad\qquad
 -E k_{\rm n}^2u_{\rm n} +F_{\rm n},
 \end{array}
\label{Shell_eq}
\end{equation}
where  $\epsilon$ is a free parameter,  $^*$ is the complex conjugate.

In the inviscid limit ($E\to 0$) and free forcing ($F=0$)
  equation (\ref{Shell_eq})
 has  two integrals.
  The first integral is the kinetic energy
\begin{equation}
{E_k}=\sum\limits_n \vert u_n\vert ^2.
\label{energ}
\end{equation}
The other integral is  the so-called generalized enstrophy
\begin{equation}
\widehat{H}=\sum\limits_n [{\rm sgn}(\epsilon -1)]^n k_{\rm n}^{\alpha(\epsilon)}\vert u_n\vert ^2,
\label{gener}
\end{equation}
with $\alpha(\epsilon)=-{\rm log}_2(\vert \epsilon -1\vert/2)$.
For $\epsilon>1$ $\widehat{H}$ is always positive
\begin{equation}
\widehat{H}=H=\sum\limits_n n k_{\rm n}^{\alpha(\epsilon)}\vert u_n\vert
^2.
\label{enstr}
\end{equation}
For $\epsilon=5/4$ its dimension coincides with dimension of the
enstrophy,  which is an integral of motion in 2D problem. That is
the reason
why  $\epsilon=5/4$  usually is associated with the
two-dimensional space. Cases with
$\epsilon<1$ correspond closer to the helicity
 integral and can have different sign.
 In the case of $\epsilon=1/2$ the dimension of $\widehat{H}$
is equal to the dimension of the hydrodynamical helicity, so this
case corresponds to  3D space. It is known that the 3D shell model
can reproduce Kolmogorov's spectrum $\sim k^{-5/3}$. The situation
in  2D space ($\epsilon = 5/4$) is more complicated. Application
of an external force at some wave number $k_f$ gives rise to two
spectral regimes: for $k>k_f$ the direct spectrum for the
enstrophy,  and  the  spectrum for the inverse energy cascade with
the slope similar to Kolmogorov's $-5/3$ for $k<k_f$. The slope of
the inverse cascade spectrum for enstrophy
 depends on $\epsilon$ and can change its value  from $-3$
for $\epsilon = 5/4$ to $-5/3$ for $\epsilon \approx 10$ \cite{FR95}.
 Here and after we will consider only the direct cascades in the 3D case with $\epsilon=1/2$.

%

To introduce effect of rotation in the GOY model we propose that
the Coriolis force can be written in the form:
\begin{equation}
{F_{\rm
c}}=-i C_{\rm r} u,
\label{Coriolis}
\end{equation}
 where the constant real coefficient $C_{\rm r}$ is introduced
for convenience.
  It is easy to see that the work of the Coriolis
  force (\ref{Coriolis}) is zero ($u_n^*{F_{\rm c}}_{\rm n}+
  u_{\rm n}{F_{\rm c}}_{\rm n}^*=0$)
  and no additional energy is introduced into the system (\ref{Shell_eq}) at any scale.
 This property of the force (\ref{Coriolis}) coincides with the property
 of the Coriolis force in 3D space (\ref{Nav}).

Having in mind that in derivation of the shell model
equations
 (\ref{Shell_eq})   all external potential forces, as well
as pressure,  were already excluded using condition of
incompressibility ($\nabla\cdot {\bf V}=0$), $F_{\rm c}$ corresponds
to the curl part of the Coriolis force. We will
return to this point when the case with the several forces will be
considered in section \ref{forcing}.

In the following sections  we
will consider the  applications of the shell model
(\ref{Shell_eq}, \ref{Coriolis}) to the free decay turbulence
(section \ref{free}) and to
 the regime with the external forcing (section  \ref{forcing}).
 Special
attention  will be paid to the helicity generation produced
in such models (section \ref{hel}).

\section{Free decay hydrodynamic turbulence}\label{free}
\begin{figure}\hskip 1cm
\epsfxsize=40cm\epsffile[150 18 1200 700]{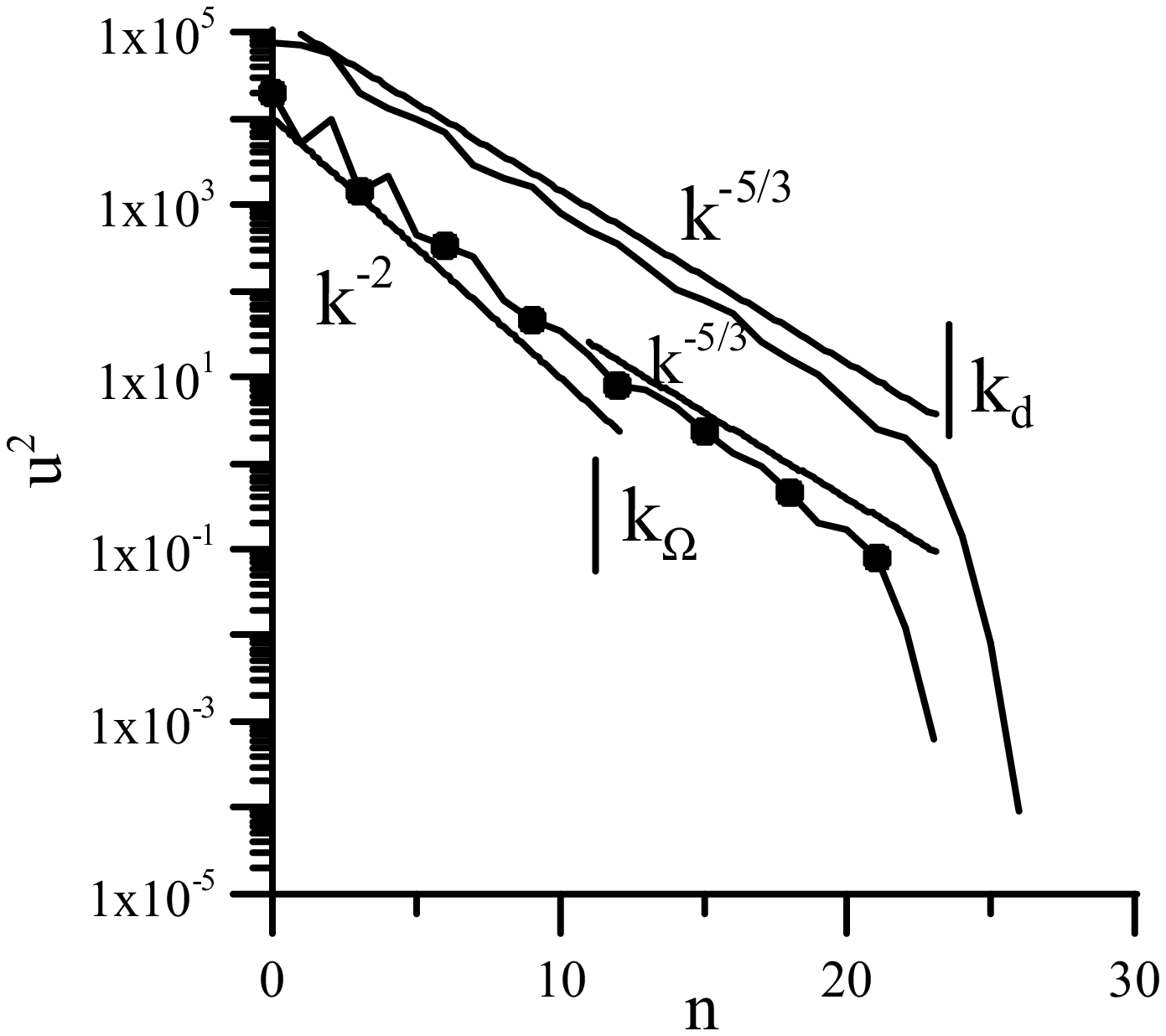}\vskip -8cm
\caption{Spectra of the free decay hydrodynamical turbulence,
$E=10^{-10},$ $R_o=10^{-3}$. Solid line corresponds to the regime
without rotation,  $C_{\rm r}=0$,  and line with circles to the regime with
rotation,
 $C_{\rm r}=1$.} \label{fig1a}
\end{figure}

\begin{figure}\hskip 1cm
\epsfxsize=40cm\epsffile[150 18 1200 700]{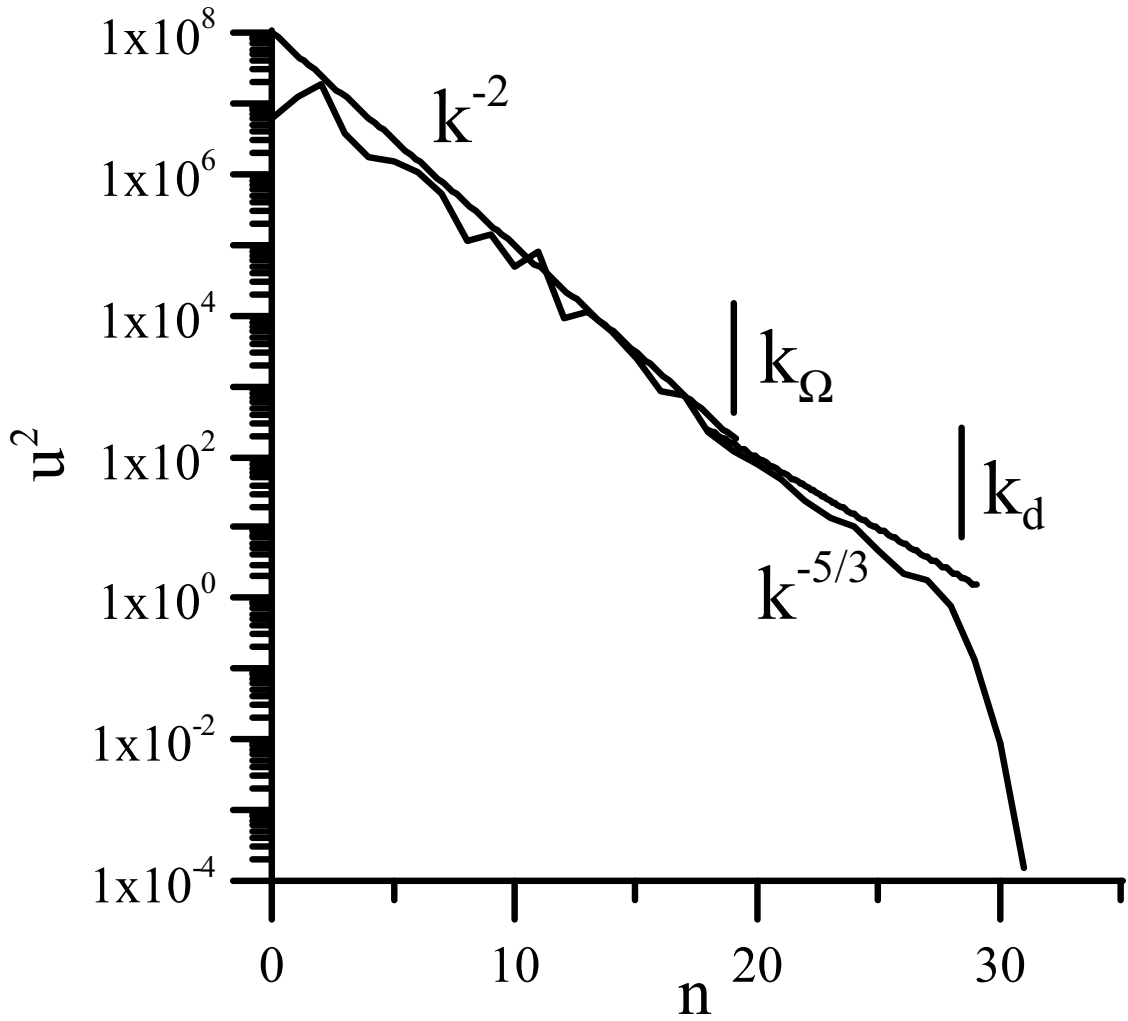}\vskip -8cm
\caption{Spectra of the free decay rotating    turbulence,
$E=10^{-15},$ $R_o=4\cdot 10^{-7}$,  $C_{\rm r}=1$.} \label{fig1b}
\end{figure}

To analize behaviour of the hydrodynamical turbulence without
forcing we start from the free decay regime ($f=0$).
After some intermediate regime for the case without rotation
($F_{\rm c}=0$)  the Kolmogorov's spectrum ($-5/3$) recovers, see
figures \ref{fig1a}, \ref{fig1b}. The sharp breaks in the spectra
at the large wave numbers correspond to the Kolmogorov's diffusion
scale $k_{\rm d}$. The estimate based on  the balance of the inertial
 $ R_{\rm o}k_{\rm n} u_{\rm n}^2$ and diffusion
$ Ek_{\rm n}^2 u_{\rm n}$ terms leads to $k_{\rm d}\sim {R_{\rm
o}\over E} u_{\rm n}$.

As was predicted in \cite{Zeman}, \cite{Zhou},
 introduction of rotation ($F_{\rm c}\ne 0$) gives rise to a
new time scale $\tau_{\rm d}\sim \Omega^{-1}$. For $R_{\rm o}\ll
1$,  $\tau_{\rm d}$ is already shorter then the characteristic time in
the Kolmogorov's regime ${\tau_{\rm d}}_{\rm n}\sim k_{\rm n
}^{-2/3}$.
 A  simple dimensional analysis  leads to the
 estimate of the rate of the energy dissipation
 \begin{equation}
  \varepsilon\sim \tau(k) k^4 E^2(k),
  \label{eps}
 \end{equation}
where $\tau$ is the characteristic time and $E(k)$ is the spectral energy density.
 In the case of the Kolmogorov's turbulence
 $\tau \sim \left(k^3E\right)^{-1/2}$ and
 \begin{equation}
  E(k)\sim \varepsilon^{2/3} k^{-5/3}.
  \label{eps1}
 \end{equation}
If the effect of  rotation is sufficient, then substitution of
$\tau\sim\Omega^{-1}$ into (\ref{eps}) leads to the rotation spectrum law \cite{Zhou}:
 \begin{equation}
  E(k)\sim \left(\Omega\varepsilon\right)^{1/2} k^{-2}.
  \label{eps2}
 \end{equation}
Starting simulations from the initial field obtained in the
non-rotating regime,   after a short time period the original
 Kolmogorov's  spectrum
splits into two parts with two different slopes, see Fig.~\ref{fig1a}, \ref{fig1b}.
The change  in the slope ($n_\Omega \in [10,19]$ and $k_{\rm n}=2^n$) corresponds to
$k_\Omega={C_\Omega^2\over R_{\rm o} u_{n_\Omega}}$, where
$C_\Omega=1.22\div 1.87$  \cite{Zhou}. This estimate can be obtained from
balancing the inertial term and the Coriolis force.
 If for the large scales the Coriolis term is
larger then the non-linear term, the spectrum decays as $\sim
k^{-2}$. In this case non-linear term do not depend on $k$ and
the Coriolis term decays as $\sim k^{-1/2}$. The further behaviour
depends on how long the spectrum is and where the Kolmogorov's
wave number $k_{\rm d}$ lies. If $k_\Omega> k_{\rm d}$, then the whole
spectrum decays as $\sim k^{-2}$. In the other case ($k_\Omega <
k_{\rm d}$) the Kolmogorov's spectrum $-5/3$ for $k>k_\Omega$
 reappears. This situation is demonstrated in the figures
\ref{fig1a}, \ref{fig1b}. As the whole kinetic energy of the
system decays in time, the $k_\Omega$ moves in to the region of the large wave numbers.
  Note
that the situation can be complicated by the high order effects,
as the  decay rates  of the both spectra are different.

\section{Convection with external forcing}\label{forcing}
To consider the evolution of the system for time periods longer
than the characteristic  decay time one needs to provide some source of
energy to the system or to introduce an external force $f$.
Similar to the case with the Coriolis force, $f$ in
(\ref{Shell_eq}) corresponds to the curl part of the force.

\begin{figure}\hskip 1cm
\epsfxsize=40cm\epsffile[150 18 1200 700]{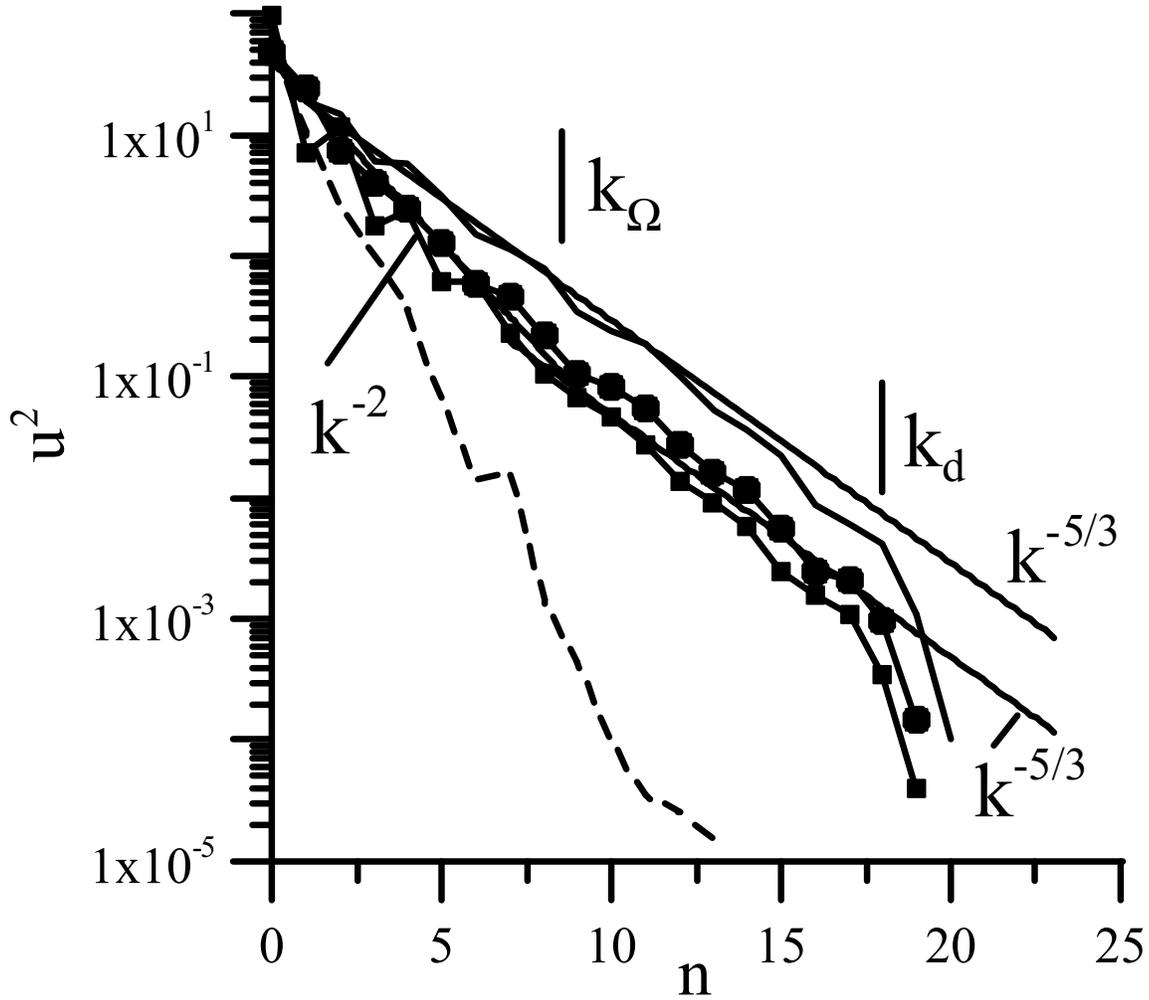}\vskip -8cm
\caption{Spectra of the  turbulence with external
forcing $f_0=10^{-2}(1+i)$, $E=10^{-10},$ $R_o=10^{-3}$. Solid line
corresponds to the regime without rotation, $C_{\rm r}=0$;  line
with circles corresponds  to the regime with rotation, $C_{\rm r}=1$ and the Coriolis
force defined by (\ref{Coriolis});
  the
dotted line to $C_{\rm r}=10$. The line with squares
 corresponds to the modified Coriolis force (\ref{Coriolis1}). } \label{fig2}
\end{figure}

The results of simulations of the system
(\ref{Shell_eq}) over the time period $t=10^3$  with a prescribed force $f=10^{-2}(1+i)$ at $k_0=1$ without rotation   are presented in
Fig.~\ref{fig2}. Starting from  an arbitrary initial velocity
field, the  system (\ref{Shell_eq}) comes to the  statistically stable state
with Kolmogorov's
energy distribution, similar to the free decay case for the moment where the energy is
 comparable.

As was already mentioned above, the direct introduction of rotation
setting $C_{\rm r}=1$ in (\ref{Coriolis})  is contradictive to the physics of the
problem and to the original equation (\ref{Nav}).
It is easy to see that for the regime of the fast
rotation ($R_{\rm o}\ll 1$) in the original Navier-Stokes equation
(\ref{Nav}), when $V<R_{\rm o}$,  the
balance between the pressure and the potential parts of the forces $f$ and $F_{\rm c}$
 holds. Such a balance of the pressure
and the Coriolis force, which takes place, e.g., in the Earth's
liquid core ($E=10^{-15}, \, R_o=4\cdot 10^{-7}$), is called the
geostrophic balance (in the present dimensionless units regime the
Earth's core situation corresponds to  $R_{\rm o}V<10^{-3}$).
Exclusion of the pressure requires exclusion  of the potential
parts of all the forces, too. Then, the remaining curl parts of
the forces are already of the same order as the inertial term and
it is these parts which are in the r.~h.~s. of (\ref{Shell_eq}).
These considerations can be formulated as follows:
\begin{equation}
\cases  {
{C_{\rm r}}_{n}=R_{\rm o} k_{\rm n}|u_{\rm n}|,
& $R_{\rm o} k_{\rm n}|u_{\rm n}|^2< |u_{\rm n}|$\cr
{C_{\rm r}}_{n}=1, & $R_{\rm o} k_{\rm n}
|u_{\rm n}|^2 > |u_{\rm n}|$.\cr }\label{cond}
\end{equation}
In other words,  condition (\ref{cond}) means that  for all
wave numbers where the Coriolis force is larger then the non-linear term it must be
compensated by the pressure, and its curl part ($F_{\rm c}$) must be of the same
order as the non-linear term. (An example of violation of the
condition (\ref{cond})  will be also considered below.)
The suggestion (\ref{cond}) is also supported by the recent paper
\cite{Constantin}, where the scaling law $\sim k^{-2}$ for
the geostrophycal balance was derived.

 The results of simulations with rotation  and  $C_{\rm r}$ defined by
 (\ref{cond}) are presented in Fig.~\ref{fig2}. As in the free
 decay case we observe two regimes with slopes $\sim k^{-2}$ for the
 small wave numbers and the Kolmogorov's regime $\sim k^{-5/3}$ for the
 wave numbers $k>k_\Omega$.  It is easy to see that in
 this model the non-linear  term  has a
 white spectrum for $k<k_\Omega$. Due to the condition (\ref{cond}),
 the curl part of Coriolis term $F_c$  has the same distribution in
 the wave space. This state of equipartition was observed after
 averaging. Analysis of the phases of the non-linear term and the
 Coriolis term reveals the existence of anticorrelation. This  is the
  reason why  the spectrum of the  rotating turbulence decreases
 faster then Kolmogorov's one. In fact, the Coriolis force
partially  locks the energy transfer from large to small scales. Moreover, the Coriolis
force blocks the applied external force, so that ${F_{\rm c}}_0=-{f_{\rm
c}}_0$
 and suppresses injection of  the
energy into the system (\ref{Shell_eq}, \ref{Coriolis}, \ref{cond}).

 To demonstrate
 this phenomenon we present calculations with $C_{\rm r}$ ten
 times larger then was predicted by (\ref{cond}) (dotted line in
 Fig.~\ref{fig2}). This regime shows very strong blocking. Our
 calculations show that the further increase of the Coriolis
 term leads to a degeneration of the spectrum at all. The
 solution then has a singular spectrum and  the whole energy is
 concentrated at the wave number $k_0$  where the force $f$ is
 applied.

\begin{figure}\vskip -1cm \hskip 1cm
\epsfxsize=40cm\epsffile[150 18 1200 700]{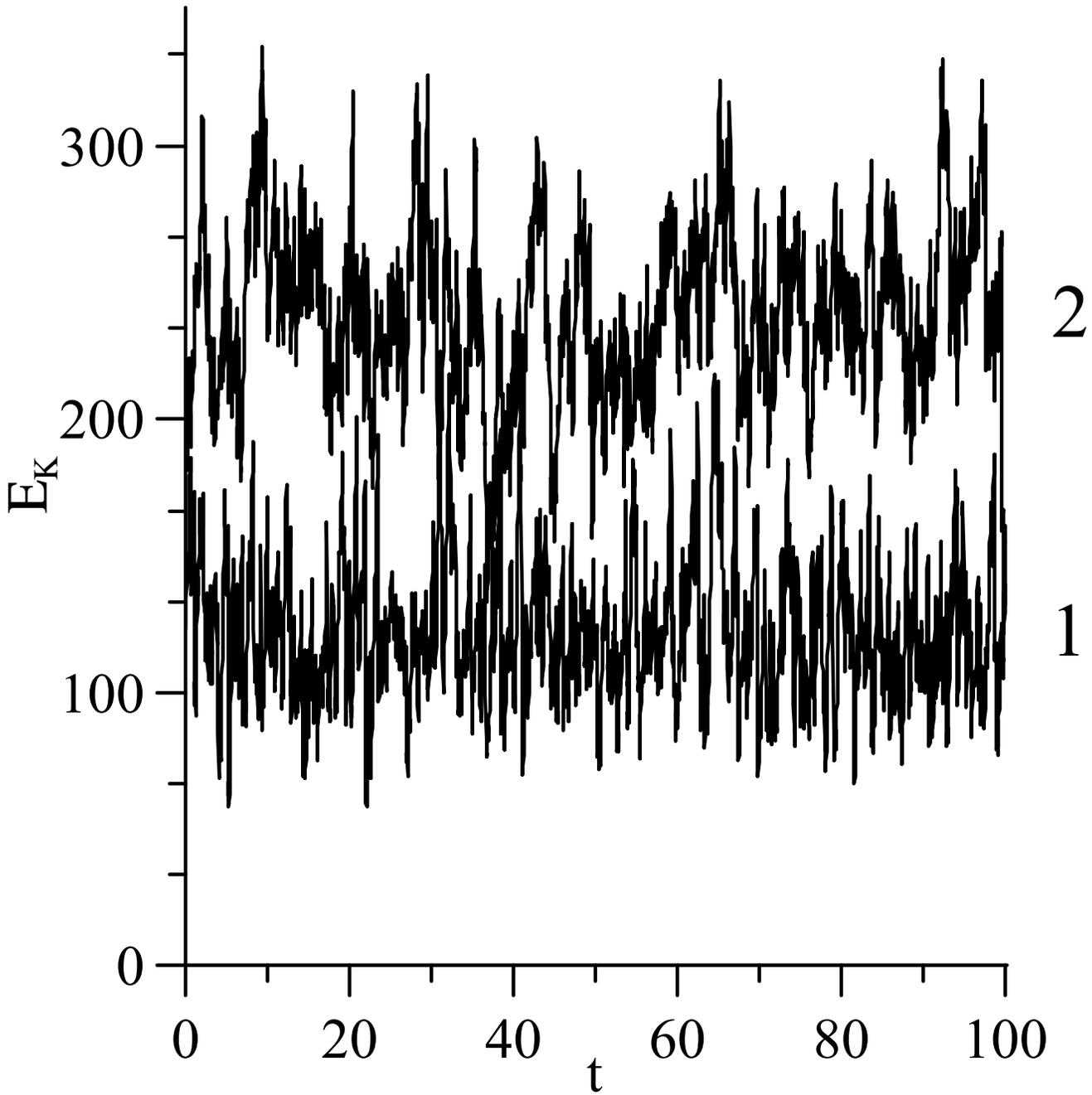}\vskip -3cm
\caption{Evolution of the kinetic energy $E_k$   for the non-rotating (1) and
 rotating (2) turbulence, $f_0=10^{-1}(1+i)/u_0^*$, $E=10^{-10},$ $R_o=10^{-3}$.
  } \label{fig2_en}
\end{figure}

We also present results of calculations with a prescribed force of
the form $f_0=0.1(1+i)/u_0^*$ which allows us to define  the exact
amount of energy injected into the system. It appears that the
introduction of rotation leads to an increase  of the total energy
 of the system  due to a decrease of the energy
 dissipation in the cascade, see  Fig.~\ref{fig2_en}.
Note that the similar effect in 3D
 numerical simulations \cite{Hossain}, where the external
force was applied at some intermediate scale,
  was caused by the inverse  cascade.

\section{Helicity generation}\label{hel}
As follows from (\ref{gener}) in 3D case ($\epsilon=1/2$) the
explicit form for helicity is
\begin{equation} H=\sum\limits_n
(-1)^n k_{\rm n}\vert u_n\vert ^2. \label{helicity}
\end{equation}
However, the direct application of formula (\ref{helicity}) for  the
estimation of helicity or the calculation of $\alpha$-effect can lead
to misleading results. Additional assumptions on the physics of the process
 are
required.  To understand the nature of the problem we will
recall some basics of the mean-field dynamo theory, where
the  models of  helicity for the rotating bodies were extensively studied.

\begin{figure}\vskip -1cm \hskip 0cm
\epsfxsize=30cm\epsffile[150 18 1200 700]{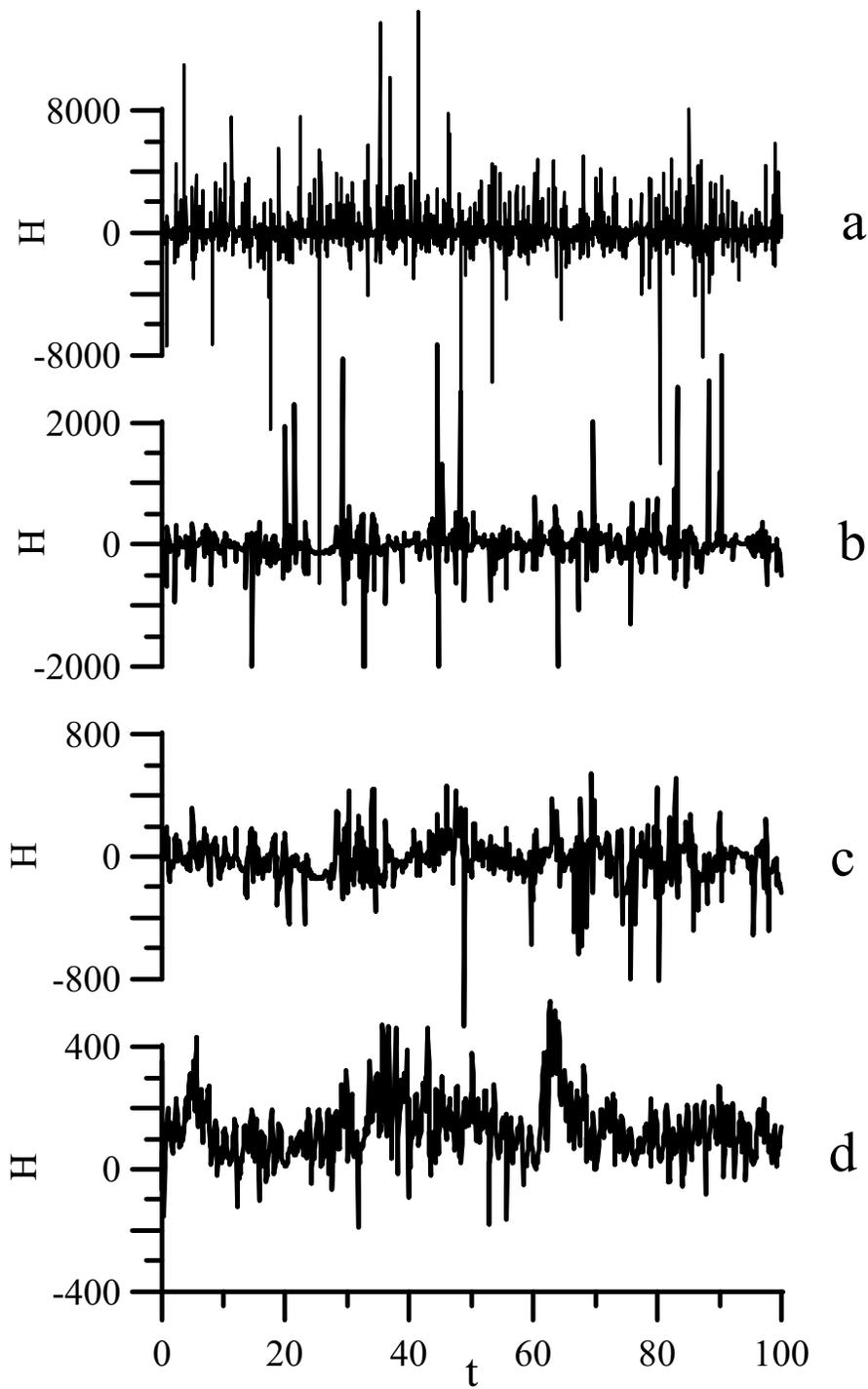}\vskip -2cm
\caption{Time evolution of  helicity H.  The curve (a) corresponds
to
 the non-rotating (Kolmogorov's) turbulence; curve (b) is the regime with
 the Coriolis force defined by (\ref{Coriolis}); case (c) is the same
  to
  (b) but with cut-off at  $k_\Omega$ (only $k \leq k_\Omega$ in
  (\ref{helicity})) and case (d) with
  the Coriolis force defined by  (\ref{Coriolis1}) and
    $k \leq k_\Omega$.} \label{fig3}
\end{figure}

According to \cite{Krause},   the mean helicity in the rotating body
  is defined  as ${\bf H}=<{\bf V}\cdot \left(\nabla\times{\bf V}\right)>$,
 provided that  the gradient of the density exists
 ($<\dots>$ means mean average). This problem
reflects the statistical inequality of the left-side and right-side  rotating curls.
In its turn,  such  inequality  is caused
by existence of the selected direction ($\bf z$),
which is related to  the global  rotation of the
considered body. In absence of rotation the generated helicity
 has stochastic temporal behavior and zero mean level. Moreover, in the case
of the Kolmogorov's turbulence with $E(k)\sim k^{-5/3}$, the helicity $H$
has an increasing spectrum $\sim k^{1/3}$ and its amplitude will be defined by the
velocity field near the Kolmogorov's cut-off wave number
$k_d$ \footnote{
Note that for the dynamo applications the maximal wave number in the sum
(\ref{helicity}) is limited by the condition $r_{\rm m} \geq 1$, where
$r_{\rm m}$ is  the
micro-scale magnetic Reynolds number.}.
The temporal behaviour of the helicity for the case of
the non-rotating homogeneous turbulence (with all other parameters are
 equal to the case with
forcing in section \ref{forcing} and summation in
(\ref{helicity}) is  over the whole range of wave numbers)  is
presented in Fig.~\ref{fig3}a. We observe  high frequency
oscillations with the negligible  mean level ($\overline{H}\sim 10^2$).
  This case can hardly
describe the helicity observed in astrophysical and geophysical  bodies
with a stable sign over  time periods much longer
then the characteristic convective times  \cite{Zeldovich}.

Application of the Coriolis force in the form (\ref{Coriolis})
does not change the situation  essentially. In  Fig.~\ref{fig3}b
we present calculations of $H$ for the previously considered
regime with rotation (Fig.~\ref{fig1a}). Even now, when the helicity spectrum is
white ($\sim k^0$) and correspondingly contributions of the high harmonics are
 smaller then in  the Kolmogorov's turbulence,
  the helicity behavior exhibits strong
oscillations with asymptotically zero mean level ($\overline{H}\sim 10^1$).

To estimate which part of helicity is produced in the rotational
part of the spectrum  ($k<k_\Omega$) we limit the
 cut-off wave number in (\ref{helicity}) to
 $k_{MAX}=k_\Omega$, see Fig.~\ref{fig3}c. This filtering
 decreases the  dispersion of the helicity, but still does not change its mean
 level ($\overline{H}\sim 10^1$).

To overcome the problem of zero mean level of helicity we have to
introduce an additional asymmetry in the system
(\ref{Shell_eq},\ref{Coriolis}).
So far in our homogeneous incompresible  shell model
the gradient of the density is
absent,  so we have  to include the  missing effect in the Coriolis force:
\begin{equation}
{F_{\rm
c}}=-i C_{\rm r} u(1+C_1(-1)^n),
\label{Coriolis1}
\end{equation}
where  $C_1$ is a real constant. The results of simulations
with $C_1=4$,   $k_{MAX}=k_\Omega$ are   presented in
Fig.~\ref{fig3}d. In this case  $\overline{H}\sim 140$ and is
comparable with the amplitude of the oscillations. The
corresponding spectrum for this regime  has the same
behavior as in the case when the Coriolis force was defined by
(\ref{Coriolis}), see Fig.~\ref{fig2}.

\section{Conclusions}
The proposed  shell model approach  description to the  rotating
 hydrodynamical turbulence demonstrates the principal possibility to
 reproduce the scaling law $\sim k^{-2}$ predicted
in \cite{Zeman}, \cite{Zhou}. This deviation from the
  homogeneous Kolmogorov's turbulence represents the reduction of the
3D  space problem toward 2D space, where the direct cascade
 of enstrophy already has
$\sim
 k^{-3}$ slope.
  In terms of the shell models one can solve the Cauchy problem
 for evaluating rotating turbulence and mimic  its dynamical
 behavior in time. As  was mentioned, the problem of rotating turbulence
  has applications to  the liquid
  core of th Earth. The estimate of the Reynolds number based on
  the west drift velocity and molecular viscosity gives $Re\sim
  10^9$, which is out of reach of any modern supercomputers in 3D
  direct numerical simulations. The situation in many other
   astrophysical bodies is even more dramatic where the estimates of the
   Reynolds number give $Re\gg 10^{10}$ \cite{Zeldovich}.
  This  Reynolds number is no problem in
  the shell model, where the wide range of scales can be
  covered by a few harmonics. That is why the proposed application of the shell model
   to the fast rotation regime  is a very
  promising step in its development.
   We also realize that the mechanism of  helicity
  generation proposed by \cite{Krause}  is not the only one and some
  forms of helicity generation  different from (\ref{Coriolis1})
  are possible. The present
 approach was dictated by its popularity in the mean-field
 dynamo model of the $\alpha$-effect.
  We plan to use the shell model developed  for the computation of the
 subgrid processes in the 3D large-scale  thermal
 convection problem similar to \cite{FRS}.

\section*{Acknowledgements}
RM  is grateful to Central Institute for Applied Mathematics (ZAM)
of Forshungszentrum in J$\rm\ddot{u}$ulich for hospitality. This
work was supported by
 the
Russian Foundation of Basic Research (grant 03-05-64074).  RM
express his gratitude to
 Peter Frick for discussions.


\pagestyle{empty}


\end{document}